\newcommand{\CLASSINPUTtoptextmargin}{1.5cm}
\newcommand{\CLASSINPUTbottomtextmargin}{1.4cm}
\newcommand{\CLASSINPUTinnersidemargin}{1.2cm}
\newcommand{\CLASSINPUToutersidemargin}{1.2cm}
\documentclass[conference]{IEEEtran}
\IEEEoverridecommandlockouts
\usepackage{cite}
\usepackage{amsmath,amssymb,amsfonts}
\usepackage{algorithmic}
\usepackage{graphicx}
\usepackage{lipsum}
\usepackage{textcomp}
\usepackage{xcolor}
\usepackage{tabularx}
\usepackage{setspace}
\usepackage{multirow}  
\usepackage{array}
\usepackage{float}
\usepackage{comment}
\usepackage[hidelinks]{hyperref}
\usepackage{enumitem}
\usepackage{xcolor}
\usepackage{soul}

\def\BibTeX{{\rm B\kern-.05em{\sc i\kern-.025em b}\kern-.08em
    T\kern-.1667em\lower.7ex\hbox{E}\kern-.125emX}}
\begin{document}

\title{Learning-based Lossless Event Data Compression
\thanks{This work was supported by RayShaper SA, Valais, Switzerland, through the Project entitled Event Aware Sensor Compression, and by the Fundação para a Ciência e a Tecnologia (FCT), Portugal, through the Project entitled Deep Compression: Emerging Paradigm for Image Coding under Grant PTDC/EEI-COM/7775/2020.}
\vspace{-6pt}
}

\author{
   \IEEEauthorblockN{Ahmadreza Sezavar\IEEEauthorrefmark{1}, Catarina Brites\IEEEauthorrefmark{2}, Jo\~{a}o Ascenso\IEEEauthorrefmark{3}}
   \IEEEauthorblockA{\IEEEauthorrefmark{1}\IEEEauthorrefmark{3}Instituto Superior Técnico,\IEEEauthorrefmark{2}Instituto Universitário de Lisboa (ISCTE-IUL), \IEEEauthorrefmark{1}\IEEEauthorrefmark{2}\IEEEauthorrefmark{3}Instituto de Telecomunicações}
    \IEEEauthorblockA{\IEEEauthorrefmark{1}ahmadreza.sezavar@lx.it.pt,\IEEEauthorrefmark{2}catarina.brites@lx.it.pt,\IEEEauthorrefmark{3}joao.ascenso@lx.it.pt}
\vspace{-25pt}
}

\maketitle

\begin{abstract}
Emerging event cameras acquire visual information by detecting time domain brightness changes asynchronously at the pixel level and, unlike conventional cameras, are able to provide high temporal resolution, very high dynamic range, low latency, and low power consumption. Considering the huge amount of data involved, efficient compression solutions are very much needed. In this context, this paper presents a novel deep-learning-based lossless event data compression scheme based on octree partitioning and a learned hyperprior model. The proposed method arranges the event stream as a 3D volume and employs an octree structure for adaptive partitioning. A deep neural network-based entropy model, using a hyperprior, is then applied. Experimental results demonstrate that the proposed method outperforms traditional lossless data compression techniques in terms of compression ratio and bits per event. 
\end{abstract}


\vspace{-12pt}
\section{Introduction}

Event cameras, also known as neuromorphic cameras or dynamic vision sensor (DVS) cameras, are currently attracting a lot of attention by the research community \cite{gallego2022event}. Unlike conventional cameras that measure absolute brightness at fixed time intervals, event cameras have a novel type of vision sensor that measures at the pixel level, in an asynchronous and independent way, brightness changes in the time domain, also known as \textit{temporal contrast}. This new way of acquiring visual information allows event cameras to achieve high temporal resolution, very high dynamic range, low latency and low power consumption. These features make event cameras attractive for different types of applications. For instance, in robotics event cameras may enable a more precise object tracking and collision avoidance while in autonomous driving they may allow real-time obstacle detection and high-speed motion estimation; additionally, the way event cameras acquire visual information allows tracking fast-moving objects without suffering motion blur, an inherent limitation of conventional cameras. 

An event camera produces a sequence of events, or an \textit{event stream}, as response to detected brightness changes in the time domain, where each event is typically represented by the 4-tuple ($x$, $y$, $t$, $p$). This 4-tuple representation encloses the basic components describing an event: the spatial coordinates $x$, $y$ (within the sensor) where the event occurred, the time $t$ at which the event occurred (also known as \textit{timestamp}) and the event polarity $p$ (+/-1) indicating the brightness change direction (increase or decrease) that triggered the event. An event stream is typically characterized by high temporal correlation while, in the spatial dimension, it is often sparse, since only pixels corresponding to moving areas and/or illumination changing areas trigger events. Thus, compression is crucial for an efficient storage, transmission, and processing of event data. However, the inherent spatial sparsity of event data, coupled with the need to maintain the spatial and temporal relationships between events, poses a challenge for achieving high compression. Recently, the JPEG standardization group has launched an exploration activity on event-based vision, called JPEG XE \cite{JPEG-XE}, precisely acknowledging the importance of developing efficient event data compression schemes. 

Lossless compression has currently been receiving more attention from the research community and has also been adopted by the JPEG XE Common Test Conditions (CTC) \cite{JPEG-Data}, which reinforces its practical importance. Lossless compression schemes exploit the inherent patterns and redundancies within the event stream, such as temporal, spatial and polarity data distributions, to encode the event data in a more compact form while allowing to obtain a perfect event data reconstruction from the compressed bitstream, without any precision loss in event representation components. This might be a rather important feature for machine vision tasks where most (if not all) the events are considered relevant to achieve high performance.

In this context, this paper proposes a novel deep-learning-based lossless event data compression scheme. The idea is to arrange the event stream as a 3D volume in $x$, $y$, and $t$ dimensions and then use an octree structure to adaptively partition the 3D volume (of events). The binary representation of the octree structure, which corresponds to a denser representation of the 3D event data volume, is then entropy coded with a learning-based entropy model. One of the main contributions of this work lies on the usage of a deep neural network to obtain the probability model of a hyperprior-based arithmetic coder. Experimental results demonstrate that the proposed solution outperforms traditional lossless data compression techniques both in terms of bits per event and compression ratio.

The rest of the paper is structured as follows: Section \ref{sec:related} briefly reviews background work on lossless event data compression, while the proposed compression method is described in Section \ref{sec:proposed}. Performance evaluation is presented and analysed in Section \ref{sec:performance}, and finally, Section \ref{sec:conclusions} concludes the paper.
\section{Background Work}\label{sec:related}
In the past years, several contributions have been made in the area of lossless event data compression. The work by Bi \textit{et al.} (2018) \cite{bi2018spike} represents one of the first and more important steps toward developing effective lossless compression schemes for event data generated by DVS. By analyzing the event pixel firing mechanism and inherent redundancies in DVS event data, a cube-based coding framework is proposed, which organizes the event sequence as a 3D space-time volume and leverages adaptive partitioning strategies to exploit the spatio-temporal correlations. The proposed encoding procedure performs independent encoding of (event) location, timestamp, and polarity data and has two prediction modes to address different event spatial distributions, the so-called address-prior mode and time-prior mode. 
\par Another strategy adopted in the literature to compress event data consists in performing temporal aggregation prior to compression; it is worth noting that temporal aggregation may introduce precision loss in the timestamp and/or polarity components of the events. In this context, events occurring within a fixed time interval at each pixel location are first aggregated or accumulated, creating a so-called (2D) event frame (EF), and the resulting EFs are then compressed \cite{schiopu2022lossless, khan2020time}. In \cite{schiopu2022lossless} Schiopu and Bilcu proposed a lossless (or lossy, if temporal aggregation is considered as part of the compression framework) compression method for encoding EFs. The proposed approach represents a group of EFs as a pair of an Event Map Image (EMI) and a polarity vector. The EMI is then represented by a binary map, that signals the spatial positions with at least one event, the number of events at each signaled position in the binary map, and the EF indices, that indicate the position(s) of the EFs associated to the events signaled within the group of EFs. While the binary map and the number of events are encoded with template context modeling, the EF indices are encoded with adaptive Markov modeling. Following a similar approach, Khan \textit{et al.} \cite{khan2020time} proposed a lossless coding strategy that leverages time aggregation and standard HEVC video coding. The proposed method temporally aggregates events according to their polarity, creating two polarity-based EFs, where each pixel location is represented by a histogram count. Next, the polarity-based EFs are concatenated before being HEVC lossless coded. In general, compression solutions based on time aggregation techniques can achieve high compression ratios, notably with large aggregation time intervals, but the loss of temporal and/or polarity precision (and event ordering) may compromise the application efficiency and introduce unacceptable delays.
\par More recently, Martini \textit{et al.} (2022) \cite{martini2022lossless} and Huang \textit{et al.} (2023) \cite{huang2023event} proposed two similar lossless compression schemes where event streams are treated as 3D point clouds. While in \cite{martini2022lossless} one 3D point cloud for each polarity is created, in \cite{huang2023event} multiple sets of two 3D point clouds are created. In both compression schemes, each 3D point cloud is then compressed using the standard MPEG Geometry-based Point Cloud Compression (G-PCC) solution.
\section{Learning-based Lossless Event Compression}\label{sec:proposed}

The proposed learning-based lossless event data compression framework, hereinafter refer to as LLEC framework, lies on two key lossless operations: i) adaptive partitioning of the 3D space-time volume of the input event data with an octree data structure; and ii) entropy coding with a learning-based entropy model, obtained from an auxiliary latent representation (hyperprior).

An octree is suited for sparse data representations like event data, as it can efficiently represent the spatial and temporal distribution of events by only subdividing regions with events, avoiding the need to represent regions without any events. The octree structure creation process begins by partitioning the spatio-temporal volume ($x$, $y$, $t$) into eight equal-sized cubes (or \textit{voxels}), each representing a given spatial region and time interval; a voxel can be either empty, i.e., containing no events, or non-empty, i.e., containing one or more events. Each non-empty voxel is then recursively divided into eight smaller (equal-sized) children voxels. The occupancy symbols of the voxels (or octree nodes) can be represented as a sequence of 8 (occupancy) binary values, where '1' represents a non-empty voxel and '0' represents an empty voxel. This binary representation of the octree structure can be lossy or lossless depending on the maximum depth in the octree structure (i.e. the maximum number of octree levels), which defines the maximum number of times the recursive process can be performed.

It is well known that entropy coding techniques, such as arithmetic coding, are lossless and can efficiently compress data when provided with an accurate probability model. In this work, it is proposed to losslessly compress the octree structure (occupancy information) using a learned hyperprior-based entropy coding model. This model statistically characterizes the source and can be obtained for each set of symbols (occupancy bytes) using side information, i.e., the hyperprior. The use of the hyperprior allows to continuously update the probability model based on the encountered symbols during encoding. This online updating strategy enables the proposed LLEC codec to efficiently capture changing data statistics and patterns, ultimately leading to more efficient compression. Moreover, the proposed learned hyperprior-based entropy model exploits the statistical dependencies in the source to learn the probability distribution of the occupancy information, allowing it to achieve much higher compression ratios compared to other alternatives, such as those that rely on the previously decoded nodes and/or the neighboring nodes at the same octree level.

\subsection{Architecture and Walkthrough}

\begin{figure}[h]
\includegraphics[width=8.5cm]{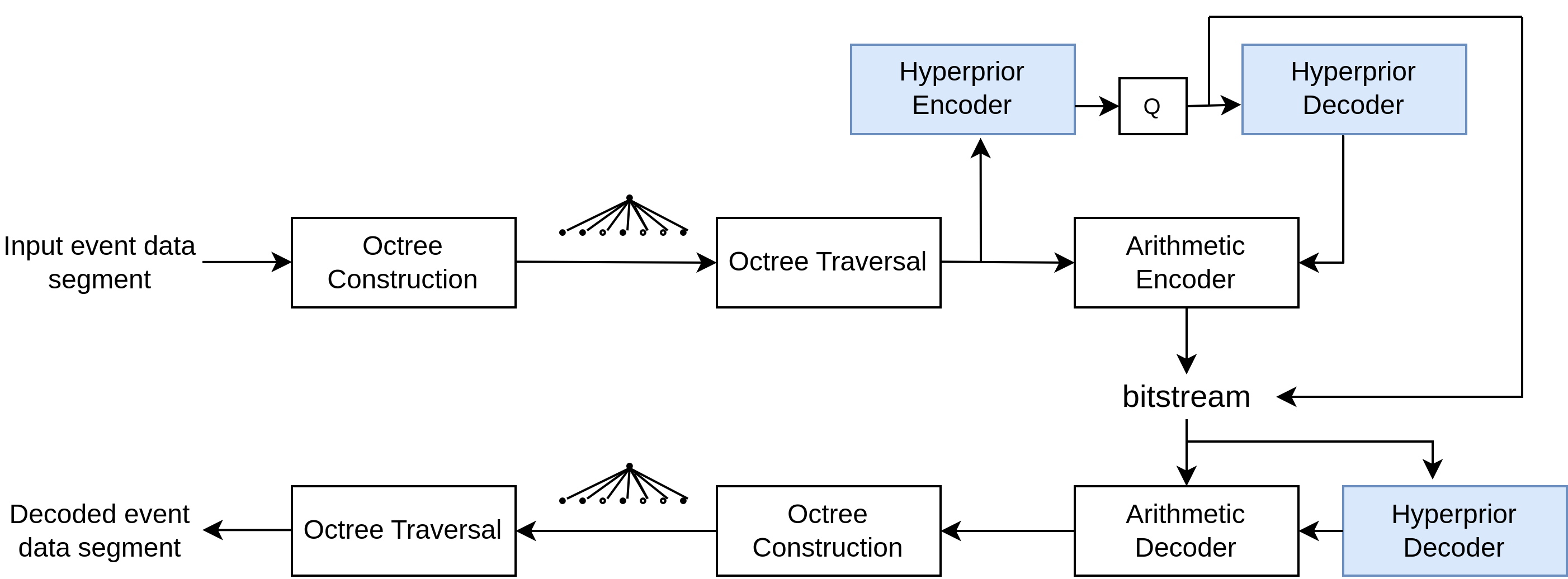}
\centering
\caption{Architecture of the proposed LLEC solution.}
\label{fig:Main_arch}
\vspace{-6pt}
\end{figure}

Figure \ref{fig:Main_arch} illustrates the proposed LLEC architecture.

Before applying the proposed LLEC solution, the input event stream is pre-processed as follows: the events of the input event stream are first aggregated, according to their polarity, into two sub-streams, one for positive $S^{+}$ and the other for negative $S^{-}$ polarity. Then, every sub-stream is temporally partitioned into multiple segments with a fixed time interval length $T_s$, $S_i^{+}, S_i^{-}$, where $i$ identifies the segment. Finally, the timestamp $t$ of each event in a segment ($S_i^{+}, S_i^{-}$) is subtracted by the minimum timestamp present in that segment, yielding to timestamp values in the range $[0 \ldots T_s]$ within each segment ($S_i^{+}, S_i^{-}$). Each pre-processed segment, independently of the polarity, is then compressed with the proposed LLEC solution following the steps below:
\begin{enumerate}[leftmargin=*]
    \item \textbf{Octree Construction}: An octree is created from the input pre-processed data $S_i^{+}, S_i^{-}$ (encoder side) or from the decoded occupancy bytes (decoder side). The depth of the octree is determined by $D=log_2(\max(x_{\text{max}},y_{\text{max}},t_{\text{max}}))$, where $x_{\text{max}}$, $y_{\text{max}}$, $t_{\text{max}}$ is the maximum value that can be assumed by spatial coordinates $x$ and $y$, and the timestamp, respectively. The computed depth $D$ guarantees that this process is lossless, i.e. each leaf node characterizes a spatio-temporal location without any ambiguity.
    \item \textbf{Octree Traversal}: At the encoder, the octree is traversed, and the occupancy bytes at each level are obtained. At the decoder, the octree is traversed until all the leaf nodes are reached, to obtain the decoded event data, which is identical to the input data. At the encoder, the octree occupancy bytes are partitioned into a set of tiles of size $N=512$ for further processing.
    \item \textbf{Hyperprior Encoder}: The hyperprior encoder is responsible to obtain a compact latent code $z$ of size $M$ for every tile of occupancies. The latent code $z$ captures the spatial dependencies among the elements of the occupancies tile and is transmitted to the decoder (after quantization) as side information; it was experimentally found that $M=8$ provides a good tradeoff between latent code compactness and overall codec performance.
    \item \textbf{Quantization}: The scalar uniform quantization approach described in Section \ref{qunat} is applied to the latent code produced by the hyperprior encoder. The quantized latent symbols $z'$ are transmitted to the decoder side without being entropy coded since the (quantized) latent code is rather compact and consumes a small amount of bitrate.
    \item \textbf{Hyperprior Decoder}: The hyperprior decoder produces the probability model of the occupancy bytes from the quantized latent symbols for every single occupancy tile. This operation is performed at both the encoder and decoder since the same probability model must be obtained at both sides. 
    \item \textbf{Arithmetic Encoder/Decoder}: An arithmetic encoder/decoder with the probability model computed in the previous step is used to code the occupancy tiles. The arithmetic codec was implemented with the torchac library \cite{torchac}\cite{mentzer2019practical}, which allows to encode, produce, parse, and decode the bitstream given the cumulative distribution of the probability model.
\end{enumerate}

\begin{figure}[h]
\includegraphics[scale=0.75]{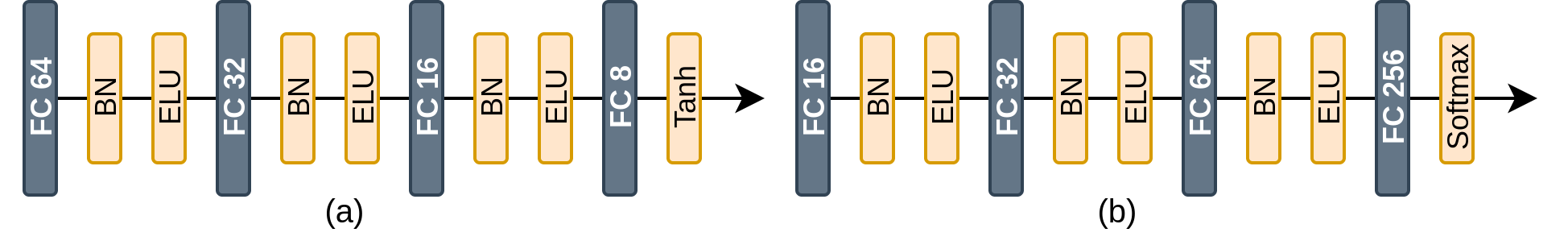}
\centering
\caption{Proposed Hyperprior network architecture: (a) encoder, (b) decoder.}
\label{fig:network}
\vspace{-10pt}
\end{figure}

\subsection{Proposed Hyperprior Network Architecture}
\par The proposed hyperprior network architecture follows an auto-encoder structure and is designed to process and transform the occupancy tiles of size $N$ through multiple layers of a neural network, capturing complex relationships within the input data. Both the hyperprior encoder (Figure \ref{fig:network}a) and hyperprior decoder (Figure \ref{fig:network}b) consist of a series of fully connected (FC) layers, followed by batch normalization (BN) and exponential linear unit (ELU) activation layers. At the hyperprior encoder, a hyperbolic tangent (tanh) activation function is employed to generate the latent code elements within a range $[-1, 1]$. At the hyperprior decoder, to produce the probability distribution model, a softmax activation function is applied as the final layer.

\subsection{Quantization}
\label{qunat}
\par The scalar quantization approach proposed in \cite{mentzer2019practical} is used to quantize the output $z$ of the hyperprior encoder. Given a set of levels $L = \{l_1, l_2, ..., l_L\} \subset \mathbb{R}$, each latent code element $z_i$ is quantized by assigning it to the nearest neighbor quantization level as defined in (\ref{equ:quant1}).
\begin{equation}\label{equ:quant1}
z_i' = Q(z_i) := \arg\min_j | z_i - l_j |
\end{equation}
The quantization operation is inherently non-differentiable, which presents a challenge to the conventional gradient-based optimization training method, as it requires all operations to be differentiable. To address this issue and facilitate the computation of gradients during the backward pass, the differentiable soft quantization approach, as defined in (\ref{equ:quant2}), is employed.
\begin{equation}\label{equ:quant2}
\hat{Q}(z_i) = \sum_{j=1}^{L} \frac{\exp(-\sigma_{q} | z_i - l_{j} |)}{\sum_{l=1}^{L} \exp(-\sigma_{q} | z_i - l_{l} |)} l_{j}
\end{equation}
In (\ref{equ:quant2}), $\sigma_q$ is a hyperparameter controlling the "softness" of the procedure. Here, the number of levels $L$ is set to 64, with the levels evenly spaced within the range $[-1, 1]$, and $\sigma_q$ is set to 2.

\section{Performance Evaluation}\label{sec:performance}
This section presents the performance evaluation of the proposed LLEC solution. In addition to the experimental results obtained (and respective analysis), this section also includes a brief description of the test material, performance metrics and training procedure adopted.

\subsection{Test Material and Benchmarks}\label{sec:tmaterial}

\par The JPEG XE reference dataset adopted by the JPEG XE CTC \cite{JPEG-Data} includes a comprehensive collection of raw event sequences, representing a wide variety of use cases, environmental conditions, durations, sensors, and sensor configurations, and it is encoded in the EVT2 format, with events sorted in chronological order based on their timestamp. In this work, almost all JPEG XE sequences were chosen for training, validation and test, and some sequences from \cite{psee-seq} were also included to have a larger test set. The selected test sequences encompass a wide range of use cases, sensor resolutions, and varying levels of activity, measured in terms of the event rate (in million events per second - Mev/s); more details on the selected sequences can be found in Table~\ref{Table:allseq}.
\par As adopted by the JPEG XE CTC for event compression \cite{JPEG-Data}, the lossless data codecs lz4 \cite{lz4}, bzip2 \cite{bzip2}, and 7z \cite{7zip} have also been used in this work to provide a reference point (anchor) for performance comparison. All experiments have been conducted over the entire sequences duration.

\vspace{-6pt}
\begin{table}[ht]
    \scriptsize
    \centering
    \caption{Event sequences selected for training, validation and test.}
    \begin{tabular}{lccc}
        \hline
        Sequence name & Spatial Res. & Duration (s) & Event-rate (Mev/s) \\
        \hline
        \multicolumn{4}{c}{\textsc{Training Sequences}} \\
        \hline
        Industrial\_spray & $640\times480$ & 1.6 & 1.30 \\
        Surveillance\_startracking & $1280\times720$ & 12.3 & 1.84 \\
        Deblur\_street & $1280\times720$ & 2.4 & 64.6 \\
        \hline
        \multicolumn{4}{c}{\textsc{Validation Sequences}} \\
        \hline
        Depthsensing\_highspeedlaser & $1280\times720$ & 3.0 & 38.8 \\
        Eyetracking\_right & $1280\times720$ & 9.9 & 5.75 \\
        \hline
        \multicolumn{4}{c}{\textsc{Test Sequences}} \\
        \hline
        Localization\_cube & $640\times480$ & 10.0 & 8.29 \\
        Activemarkers\_handheld & $1280\times720$ & 10.1 & 0.60 \\
        Industrial\_counting & $640\times480$ & 6.3 & 0.73 \\
        Industrial\_fluidflow & $1280\times720$ & 5.6 & 1.88 \\
        \textit{Spinner} & $624\times477$  &  5 &    10.83   \\
        \textit{Hand\_spinner} &   $639\times479$ &  5 &   2.39  \\
        \textit{80\_balls} &   $639\times479$ & 6.3  &   0.72 \\
        \textit{195\_falling\_particles} & $633\times479$  & 0.032 &   632.52 \\
        \textit{Monitoring\_40\_50hz }&  $639\times479$ & 6  &  6.77  \\ 
        \textit{Traffic\_monitoring} & $640\times480$ & 28 & 0.64\\
        \textit{Cube}&  $640\times480$ & 17.7  &  7.00  \\ 
        \hline
    \label{Table:allseq}
    \end{tabular}
\vspace{-20pt}
\end{table}

\subsection{Performance Metrics}
For performance evaluation, two metrics, also adopted by the JPEG XE CTC \cite{JPEG-Data}, are used: the compression ratio (CR) and the average compressed event size in bits (S). The CR metric measures the reduction in size of the compressed bitstream generated by the proposed LLEC solution (in bits) compared to the size (in bits) of the input sequence represented in the EVT2 format, as defined in (\ref{equ:cr}). It is important to note that the compressed rate includes both the bits necessary to transmit the hyperprior and the arithmetic coded symbols.
\begin{equation}\label{equ:cr}
CR = \frac{\text{size}(\text{input sequence})}{\text{size}(\text{compressed bitstream})}
\end{equation}
The S metric is also a way to represent compression efficiency. S quantifies the average number of bits used by the proposed LLEC solution to represent an event and it is computed as the ratio between the compressed bitstream size in bits and the total number of events in the input sequence, expressed in bits/event, according to (\ref{equ:s}).
\begin{equation} \label{equ:s}
S = \frac{\text{size}(\text{compressed bitstream})}{\text{number of input events}} 
\end{equation}

\subsection{Training Procedure}
In the pre-processing step, it was set the time interval $T_s=2048$ and $T_s=1024$ for sequences with spatial resolution of $1280\times780$ and $640\times480$, respectively; these $T_s$ values are also applied during inference. In the training procedure, validation sequences are used to fine-tune the hyperparameters, such as the number of layers, the batch size, and the learning rate. The Adam optimizer \cite{kingma2015adam} is used, with a learning rate of $10^{-4}$. A scheduler with a decay rate of 0.1 every 5 epochs is also employed. The proposed hyperprior features a latent code size of 8 and a batch size of 512, as these values have proven to be the best experimentally. Before the training begins, segments are randomly selected from all training event sequences until 110K tiles are obtained. Training occurs over 100 epochs but, to prevent overfitting, an early-stopping mechanism with a patience of 10 epochs is implemented, monitoring validation loss for improvements. The hyperprior network training is performed on a machine equipped with an NVIDIA GeForce RTX 3090 GPU. Regarding the loss function, the objective of the proposed learned hyperprior-based entropy model is to minimize the expected length of the bitstream (rate) according to:
\begin{equation}\label{equ:loss}
R = \mathbb{E}_{x \sim p(x)} \left[-\log_2 \hat{p}\left(x\right)\right] 
\end{equation}  
where $x$ is the input occupancy tile, $p(x)$ is the unknown distribution of (input) occupancy tiles and $\hat{p}(x)$ is the discrete entropy model estimated by the hyperprior decoder. Equation (\ref{equ:loss}) represents the cross-entropy between the marginal distribution of the latent code and the learned entropy model, and is used in the training as the loss function; the cross-entropy, or loss, is minimized when the two distributions are identical.

\subsection{Experimental Results and Analysis}
Table \ref{table:CR_Metric} and Table \ref{table:S_Metric} show the experimental results obtained with the proposed LLEC solution and benchmarks for the CR and S metrics, respectively, for the test sequences listed in Table \ref{Table:allseq}; these sequences represent an activity variety (see Section \ref{sec:tmaterial}) that is important to evaluate the LLEC performance under different data throughput conditions and, thus, to obtain enough representative performance results. From Table \ref{table:CR_Metric} and Table \ref{table:S_Metric}, the following conclusions can be taken:

\begin{itemize}[leftmargin=*]
\item \textbf{Compression Ratio (CR)}: As it can be seen from Table \ref{table:CR_Metric}, the proposed LLEC solution outperforms all benchmark methods (lz4, bzip2, and 7z) in terms of CR for all test sequences, achieving CRs gains up to 3.5x, 2.6x and 2.1x over lz4, bzip2, and 7z, respectively; higher CR values indicate better compression performance. The CR gains and its consistency across sequences with diversified activity levels and over different benchmarks demonstrates the enhanced compression capability of the proposed LLEC solution.

\item \textbf{Average compressed event size in bits (S)}: In terms of the S metric, the proposed LLEC solution also outperforms all benchmark methods (lz4, bzip2, and 7z) for all test sequences, achieving S values significantly lower than the benchmark ones (up to 3.4x, 2.6x and 2.0x lower, resp.); lower S values indicate better compression performance. The S reduction magnitude and its consistency across sequences with diversified activity levels and over different benchmarks reinforces the improved compression capability of the proposed LLEC solution.

\item \textbf{Computational complexity of the hyperprior model}: The hyperprior model consists of an encoder and a decoder, each with distinct computational requirements. The encoder exhibits a computational complexity of 36.02 KMAC (multiply-accumulate operations) and has 35.8k parameters. In contrast, the decoder demonstrates a lower computational demand, with 19.89 KMAC and 19.66k parameters. The architecture designed balances the model's expressive power with complexity, with the encoder handling a more computationally intensive task compared to the decoder.

\end{itemize}
\vspace{-12pt}

\begin{center}
\begin{table}[ht]
\scriptsize
  \centering
  \renewcommand{\arraystretch}{1.2}
  \caption{Compression Efficiency Results for the CR Metric.}
  \begin{tabular}{|c|*{4}{c|}}
    \hline
    \multirow{2}{*}{\bfseries Sequence name} & \multicolumn{4}{c|}{\bfseries CR} \\
    \cline{2-5}
     & \textbf{Proposed LLEC} & \textbf{lz4} & \textbf{bzip2} & \textbf{7z} \\
    \hline
    Industrial\_counting &  \textbf{1.99} & 0.93 & 1.28 & 1.56 \\
    \hline
    Activemarkers\_handheld &  \textbf{2.85} & 0.90 & 1.27 & 1.86 \\
    \hline
    Localization\_cube &  \textbf{3.29} & 0.99 & 1.25 & 1.84 \\
    \hline
    Industrial\_fluidflow &  \textbf{1.37} & 0.94 & 1.10 & 1.14 \\
    \hline
    Spinner &  \textbf{3.05} & 0.99 & 1.51 & 2.59 \\
    \hline
    Hand\_spinner &  \textbf{1.73}  & 0.97 & 1.35 & 1.45 \\
    \hline
    80\_balls &  \textbf{1.99} & 0.93 & 1.28 & 1.56\\
    \hline
    195\_falling\_particles &  \textbf{2.69} & 0.99 & 1.21 & 1.45\\
    \hline
    Monitoring\_40\_50hz &  \textbf{1.78} & 0.99 & 1.26 & 1.54 \\ 
    \hline
    Traffic\_monitoring &  \textbf{1.52} & 0.97 & 1.25 & 1.35 \\
    \hline
    Cube & \textbf{2.29} & 0.99 & 1.25 & 1.84 \\
    \hline
  \end{tabular}
  \label{table:CR_Metric}
  \vspace{-30pt}
\end{table}
\end{center}

\begin{center}
\begin{table}[ht]
  \scriptsize
  \centering
  \caption{Compression Efficiency Results for the S Metric.}
  \begin{tabular}{|c|*{4}{c|}}
    \hline
    \multirow{2}{*}{\bfseries Sequence name} & \multicolumn{4}{c|}{\bfseries S} \\
    \cline{2-5}
     & \textbf{Proposed LLEC} & \textbf{lz4} & \textbf{bzip2} & \textbf{7z} \\
    \hline
    Industrial\_counting &  \textbf{16.02}  & 34.26   & 24.92  & 20.40 \\
    \hline
    Activemarkers\_handheld &  \textbf{11.19}  & 35.17  & 25.10  & 17.11 \\
    \hline
    Localization\_cube &  \textbf{9.71} & 32.24   & 25.51 & 17.31 \\
    \hline
    Industrial\_fluidflow &  \textbf{23.34} & 33.71 & 29.05 & 27.90 \\
    \hline
    Spinner &  \textbf{10.45} & 32.17 & 21.14 & 12.32 \\
    \hline
    Hand\_spinner &  \textbf{18.47}  & 32.83 & 23.58 & 22.00 \\
    \hline
    80\_balls &  \textbf{16.02} & 34.26 & 24.92 & 20.40 \\
    \hline
    195\_falling\_particles &  \textbf{11.88} & 32.00 & 26.31 & 21.95 \\
    \hline
    Monitoring\_40\_50hz &  \textbf{17.90} & 32.29 & 25.25 & 20.72 \\ 
    \hline
    Traffic\_monitoring &  \textbf{20.96} & 32.78 & 25.44 & 23.55 \\
    \hline
    Cube &  \textbf{13.97} & 32.28 & 25.58 & 17.34 \\
    \hline
  \end{tabular}
  \label{table:S_Metric}
  \vspace{-20pt}
\end{table}
\end{center}

\section{Conclusions}\label{sec:conclusions}
\par This work proposes a novel learning-based lossless event data compression framework, where events are represented with an octree data structure that is arithmetic coded with a hyperprior-driven entropy model. The proposed hyperprior network architecture encloses two neural networks following an auto-encoder structure, which allows to effectively capture the source statistics. Experimental results demonstrate that the proposed method consistently outperforms existing lossless data compression approaches for sequences with different activity levels.
\bibliographystyle{ieeetr}
\bibliography{Ref}
\end{document}